\documentclass[conference, 11pt]{IEEEtran}
\usepackage[margin=1.0in]{geometry}
\usepackage{graphicx}
\usepackage[utf8]{inputenc}
\usepackage{fontenc}
\usepackage{subcaption}
\usepackage{csquotes}
\usepackage{amsmath}
\usepackage{textcomp}
\usepackage{multirow}
\usepackage{float}
\usepackage{mathtools}
\usepackage{cite}
\usepackage{filecontents}
\usepackage{stfloats}
\usepackage{balance}
\usepackage{mathcomp}
\usepackage{spverbatim}
\usepackage{verbatimbox}
\usepackage[usenames, dvipsnames]{color}
\usepackage{algorithm2e}
\usepackage[english]{babel}
\usepackage{blindtext}
\usepackage{array}
\usepackage{MnSymbol}
\newcolumntype{L}[1]{>{\raggedright\let\newline\\\arraybackslash\hspace{0pt}}m{#1}}
\newcolumntype{C}[1]{>{\centering\let\newline\\\arraybackslash\hspace{0pt}}m{#1}}
\newcolumntype{R}[1]{>{\raggedleft\let\newline\\\arraybackslash\hspace{0pt}}m{#1}}

\usepackage[table,xcdraw]{xcolor}

\begin{document}

\title{Hardware Software Co-design framework for Data Encryption in Image Processing Systems for the Internet of Things Environment}

% \author{Kusum Lata, \IEEEmembership{Senior Member, IEEE}, Surbhi Chhabra , \IEEEmembership{Member, IEEE}, and Sandeep Saini, \IEEEmembership{Member, IEEE}}

% \author{
%     \IEEEauthorblockN{Kusum Lata, Surbhi Chhabra, and Sandeep Saini}\\
%     \IEEEauthorblockA{The LNM Institute of Information Technology, Jaipur (Rajasthan), India}}
    
\author{\IEEEauthorblockN{Kusum Lata}
\IEEEauthorblockA{The LNMIIT, Jaipur\\}
\and
\IEEEauthorblockN{Surbhi Chhabra}
\IEEEauthorblockA{The LNMIIT, Jaipur}\\
\and
\IEEEauthorblockN{Sandeep Saini}
\IEEEauthorblockA{The LNMIIT, Jaipur}
}

% \author{Kusum Lata, Surbhi Chhabra, and Sandeep Saini}

\maketitle

\begin{abstract}
Data protection is a severe constraint in the heterogeneous IoT era. This article presents a Hardware-Software Co-Simulation of AES-128 bit encryption and decryption for IoT Edge devices using the Xilinx System Generator (XSG). VHDL implementation of AES-128 bit algorithm is done with ECB and CTR mode using loop unrolled and FSM-based architecture. It is found that AES-CTR and FSM architecture performance is better than loop unrolled architecture with lesser power consumption and area. For performing the Hardware-Software Co-Simulation on Zedboard and Kintex-Ultra scale KCU105 Evaluation Platform, Xilinx Vivado 2016.2 and MATLAB 2015b is used. Hardware emulation is done for grey images successfully. To give a practical example of the usage of proposed framework, we have applied it for Biomedical Images (CTScan Image) as a case study. Security analysis in terms of the histogram, correlation, information entropy analysis, and keyspace analysis using exhaustive search and key sensitivity tests is also done to encrypt and decrypt images successfully. 
\end{abstract}

%\begin{IEEEkeywords}
%AES-128, VHDL, Image encryption, Image decryption, Hardware Co-Simulation, Internet of Thing, Image Processing, Cryptography, FPGA
%\end{IEEEkeywords}

\section{Introduction}
In the era of the heterogeneous IoT environment around us, where data transmission continues to happen every second, protecting these data is a daunting task; however, with proper cryptography, these security challenges can be alleviated. Cryptography algorithms are commonly used in applications where security is critical, such as bank services, military services, ATM cards, computer passwords, online money transactions, and e-commerce; they are also used in wireless devices \cite{toubal2020fpga}. AES crypto-algorithm is the most commonly accepted and used crypto-algorithm for providing security to IoT devices \cite{arab2019image}.  It is the most appropriate lightweight crypto primitives that help IoT devices to transfer images securely over the present heterogeneous IoT environment \cite{dhanda2020lightweight}.  On the Virtex-6  FPGA board, the AES algorithm for image encryption is implemented using Xilinx System Generator (XSG) inbuilt boxes \cite{arshad2014fpga}. Since multimedia applications affect many facets of our lives, a hardware implementation of AES-based real-time video encryption is presented \cite{kotel2014fpga}, where the video encryption is tested and shown on a DVI monitor. In crypto-module hardware implementations, side-channel attacks are a significant concern. Therefore, in \cite{8719373}, the security of the AES algorithm has been improved by obfuscating the specification, which was then used in the hardware-software co-design of the AES-128 for image processing applications. Modified AES is implemented on FPGA for medical image encryption and decryption \cite{hafsa2021fpga}. The implementation is also compared with the recent image cryptosystems, and it is claimed that their performance is more robust and efficient. Recently, a new concept of building security primitives from solar cells and sensors is also reported in \cite{9426081}, which also could be integrated with crypto-algorithms to enhance the data protection in Cyber-Physical Systems. Previously, we proposed Hardware Software Co-Simulation of AES-128 based data encryption using Xilinx ISE 14.2 and Matlab 2011a on spartan-6 FPGA board for both Grey and Colored images \cite{degada2020integrated}.  
\par The significant contributions of this article are summarized as follows:
\begin{enumerate}
 \item Hardware-Software Co-Design framework of AES-128 bit algorithm is presented. The AES algorithm is first written in VHDL and then implemented using XSG (integration of Xilinx Vivado 2016.2 and MATLAB 2015b blocks) on Zedboard (XC7Z0201CLG484C) and Kintex-Ultra scale KCU105 evaluation platform.
 \item The proposed framework is implemented for two modes of AES crypto-algorithm, i.e., ECB and CTR. Further, to implement these two modes, loop unrolled, and FSM-based optimization architectures are used.
 \item Co-design framework has been analyzed using resource utilization summary and throughput. Further, these results are compared to other state-of-the-art existing works. It is observed that the proposed CTR mode AES-128 gives significant improvement with fewer resource utilization and better efficiency. 
\item Security analysis is done for lung cancer CT scan images through Histogram, correlation, Information Entropy, Key Sensitivity Test and Differential Attack analysis. It shows that the proposed Co-design implementation is resistant to statistical and differential attacks.   
 \end{enumerate}

\section{IMPLEMENTATION OF AES-128 BIT ALGORITHM USING XSG}
This work implements a hardware-software co-design for secure image processing by exploring different modes and architectures of the AES-128 crypto-algorithm. Implementation of AES-128 algorithm is done for two modes i.e. ECB and CTR mode. To optimize the whole design loop unrolled and FSM based architectures are implemented for each mode of AES-128. Algorithms are optimized for gaining higher security with lesser resource utilization and better efficiency. Proposed platform design is emulated on Zedboard (XC7Z0201CLG484C) and Kintex-Ultra scale KCU105 evaluation platform.

\par AES-128 algorithm is implemented first at the RTL level using VHDL. Hardware-Software co-simulation is performed for grey images. To execute the co-simulation of the AES-128 crypto-module, the first Simulink model is built and then it is integrated with the XSG, Xilinx Vivado 2016.2, and MATLAB 2015b. The experiment is performed on a HOST PC having the following specifications: Intel corei7 processor, 64-bit operating system, and 16 GB RAM. 

\subsection{Hardware Co-Simulation of AES-128 bit algorithm for Grey Images: }
The co-simulation model for the grey image of size $440*123$ pixels for Zedboard (XC7Z0201CLG484C) FPGA board using XSG is shown in Figure 2. In this Simulink model, Zedboard$\_$hwcosim is the netlist generated for the FPGA board. A black box named Bit conversion having a VHDL code of converting the 8-bit output of gateway into 128 bit, as we need 128-bit block length and key length for the AES algorithm. The encryption black box having the VHDL coding of AES encryption gives the outcome as an encrypted image. The Decryption black box having the VHDL coding of AES decryption gives the result as a decrypted image shown on the video viewer block.

\begin{figure*}
\centering
\includegraphics[scale=0.8]{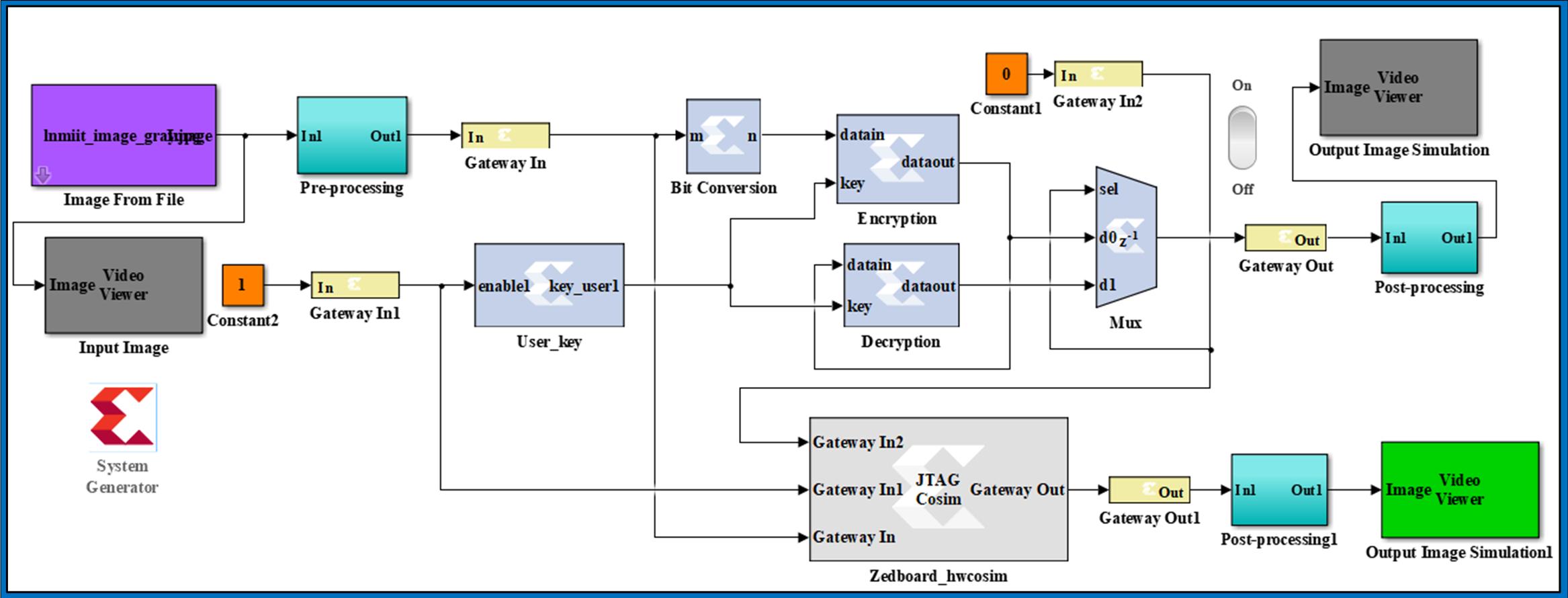}
\caption{AES-128 bit implementation on Zedboard for the grey image using XSG}
\end{figure*}

\section{Results and Analysis}
Results are analyzed in the following ways: A) VHDL implementation of AES-128 with ECB and CTR mode of operation using loop unrolled and FSM architecture on Zedboard and Kintex-Ultra scale KCU105 Evaluation Platform. Implementation results are compared with similar existing work in terms of area, power, and throughput. B) Successfully encrypted and decrypted grey and colored images and examined the statistical and differential attacks. 

\subsection{Resource Utilization and Throughput}
Table I shows the implemented architectures’ resource utilization summary for AES-128 with ECB and CTR mode of operation implemented on Zedboard and Kintex ultra-scale platform. The table shows that AES-FSM and AES-CTR architecture utilize optimum LUTs and FFs within 20\% of the available resources compared to loop unrolled AES. Throughput (T) is calculated as follows for the implemented AES-128.

\begin{equation}
\begin{multlined}
\mathrm{\text{T}}= \frac{\text{No. of processed bits} \times \text{Maximum Frequency}}{\text{Latency}}
\end{multlined}
\end{equation}

The maximum frequency for AES-ECB loop unrolled, FSM, and AES CTR is 119.35MHz, 211.8MHz, and 175.35MHz for ZedBoard FPGA. The corresponding throughput for all the architecture is 1.91Gbps, 3.39Gbps, and 2.04Gbps, respectively, as shown in Table II.In contrast, the maximum frequency comes out to be AES-ECB loop unrolled, FSM, and AES CTR are 345MHz, 245.62MHz, and 222.22MHz for Kintex Ultrascale, respectively. And the corresponding throughput for all the architecture is 5.52Gbps, 2.35Gbps, and 2.58Gbps, respectively.It can also be seen from the table that FSM based architecture of ECB mode is taking only 4058 LUTs and 0.248 W power consumption, which is much lower than the area reported by \cite{8719373},\cite{soliman2016efficient},\cite{bentoutou2020improved}. Similarly, throughput and efficiency is improved significantly with respect to the work reported by \cite{8719373} and \cite{bentoutou2020improved}, where as in \cite{soliman2016efficient} it is not reported. From the table, we can conclude that AES-ECB FSM architecture exhibits high throughput with minimum resource utilization. But, AES-ECB mode is not suitable for encrypting the image data in high-security applications; therefore, we prefer AES-CTR with optimized pipeline architecture.

\begin{table*}
\centering
\caption{Resource Utilization for AES-128 ECB Mode Loop Unrolled \& FSM and CTR Mode Architecture}
\begin{tabular}{|L {1.4cm}|C {1.2cm}|C {1.4cm}|C {1.4cm}|C {1.4cm}|C {1.3cm}|C {1.4cm}|C {1.4cm}|C {1.4cm}|}
\hline
 & \multicolumn{4}{c|}{\textbf{ZedBoard (XC7Z0201CLG484C)}} & \multicolumn{4}{c|}{\textbf{Kintex Ultrascale KCU105 Evaluation Platform}} \\ \hline
 &  & \begin{tabular}[c]{@{}c@{}}Loop \\ Unrolled\\ ECB Mode\end{tabular} & \begin{tabular}[c]{@{}c@{}}FSM\\ ECB Mode\end{tabular} & CTR Mode &  & \begin{tabular}[c]{@{}c@{}}Loop \\ Unrolled\\ ECB Mode\end{tabular} & \begin{tabular}[c]{@{}c@{}}FSM\\ ECB Mode\end{tabular} & CTR Mode \\ \hline

Resources & Available & Utilization (\%) & Utilization (\%) &Utilization (\%) & Available  &Utilization (\%) &Utilization (\%) &Utilization (\%) \\ \hline 

LUT & 53200& 31.36 &10.21 & 15.34& 242400 &6.93 & 2.24 & 3.3  \\ \hline 

%LUT & 53200 & \begin{tabular}[c]{@{}c@{}}16681 (31.36)\end{tabular} & \begin{tabular}[c]{@{}c@{}}5432 (10.21)\end{tabular} &  \begin{tabular}[c]{@{}c@{}}8160 (15.34)\end{tabular} & 242400 & \begin{tabular}[c]{@{}c@{}}16795 (6.93)\end{tabular} & \begin{tabular}[c]{@{}c@{}}5426 (2.24)\end{tabular} &  \begin{tabular}[c]{@{}c@{}}8160 (3.3)\end{tabular}\\ \hline
FF & 106400 &1.02  &1.35 &6.03 &484800 &0.23 &0.3&1.32 \\ \hline 
IO &200  &192  &198.5 & 193& 520& 73.84& 76.34 & 74.23\\ \hline 

%FF & 106400 & \begin{tabular}[c]{@{}c@{}}1088\\ (1.02 \%)\end{tabular} & \begin{tabular}[c]{@{}c@{}}1437\\ (1.35 \%)\end{tabular} & \begin{tabular}[c]{@{}c@{}}6416\\ (6.03 \%)\end{tabular} &   484800 & \begin{tabular}[c]{@{}c@{}}1091\\ (0.23 \%)\end{tabular} & \begin{tabular}[c]{@{}c@{}}1437\\ (0.3 \%)\end{tabular} & \begin{tabular}[c]{@{}c@{}}6416\\ (1.32 \%)\end{tabular} \\ \hline
%IO & 200 & \begin{tabular}[c]{@{}c@{}}384\\ (192 \%)\end{tabular} & \begin{tabular}[c]{@{}c@{}}397\\ (198.5 \%)\end{tabular} &  \begin{tabular}[c]{@{}c@{}}386\\ (193 \%)\end{tabular} & 520 & \begin{tabular}[c]{@{}c@{}}384\\ (73.84 \%)\end{tabular} & \begin{tabular}[c]{@{}c@{}}397\\ (76.35 \%)\end{tabular} & \begin{tabular}[c]{@{}c@{}}386\\ (74.23 \%)\end{tabular} \\ \hline

BUFG & 32 & 12.5 & 3.13 & 3.13 & 480 & 1.04 & 0.21 & 0.21\\ \hline
BRAM & 140 & 1.43 & - & - & 600 & 0.33 & - & - \\ \hline
MMCM & 4 & 25 & - & - & 10 & 10 & - & - \\ \hline

\end{tabular}
\end{table*}
% \begin{table}[]
% \centering
% \begin{tabular}{|c|c|c|c|c|c|}
% \hline
% Reference & Device & Area & \begin{tabular}[c]{@{}c@{}}Power\\ (W)\end{tabular} & \begin{tabular}[c]{@{}c@{}}Throughput\\ (Gbps)\end{tabular} & \begin{tabular}[c]{@{}c@{}}Efficiency\\ (Mbps)\end{tabular} \\ \hline
% 18 & XC7Z010clq225 & NA & 0.675 & NA & NA \\ \hline
% 16 & ZedBoard & 16681 & NA & 5.48 & 0.266 \\ \hline
% 8 & \begin{tabular}[c]{@{}c@{}}Artix-7\\ XC7A100T\end{tabular} & 6568 & 1.37 & 1.75 & 0.32 \\ \hline
% \begin{tabular}[c]{@{}c@{}}Loop\\ Unrolled\\ ECB\end{tabular} & ZedBoard & 16681 & 1.058 & 1.91 & 0.12 \\ \hline
% \begin{tabular}[c]{@{}c@{}}FSM\\ ECB\end{tabular} & ZedBoard & 4058 & 0.248 & 3.93 & 0.96 \\ \hline
% \begin{tabular}[c]{@{}c@{}}Loop\\ Unrolled\\ ECB\end{tabular} & \begin{tabular}[c]{@{}c@{}}Kintex\\ Ultrascale\end{tabular} & 16795 & 1.214 & 5.52 & 0.33 \\ \hline
% \begin{tabular}[c]{@{}c@{}}FSM\\ ECB\end{tabular} & \begin{tabular}[c]{@{}c@{}}Kintex\\ Ultrascale\end{tabular} & 4058 & 0.493 & 2.35 & 0.58 \\ \hline
% \begin{tabular}[c]{@{}c@{}}AES\\ CTR\end{tabular} & ZedBoard &  &  &  &  \\ \hline
% \begin{tabular}[c]{@{}c@{}}AES\\ CTR\end{tabular} & \begin{tabular}[c]{@{}c@{}}Kintex\\ Ultrascale\end{tabular} &  &  &  &  \\ \hline
% \end{tabular}
% \end{table}

\begin{table*}
\centering
\caption{Comparison with the existing work}
\begin{tabular}{|L {2.8cm}|L {3.5cm}|C {1.5cm}|C {1.5cm}|C {1.8cm}|C {1.8cm}|}
\hline
Reference & Device & Area (LUTs) & Power (W) & Throughput (Gbps) & Efficiency (Mbps) \\ \hline
\cite{8719373} & ZedBoard & 16681 & NA & 5.48 & 0.266 \\ \hline
\cite{soliman2016efficient} & Zynq-7 XC7Z010clq225 & NA & 0.675 & NA & NA \\ \hline
\cite{bentoutou2020improved} & Artix-7 XC7A100T & 6568 & 1.37 & 1.75 & 0.32 \\ \hline
Loop Unrolled ECB & ZedBoard & 16681 & 1.058 & 1.91 & 0.12 \\ \hline
FSM ECB & ZedBoard & 4058 & 0.248 & 3.93 & 0.96 \\ \hline
Loop Unrolled ECB & Kintex Ultrascale & 16795 & 1.214 & 5.52 & 0.33 \\ \hline
FSM ECB & Kintex Ultrascale & 4058 & 0.493 & 2.35 & 0.58 \\ \hline
AES CTR & ZedBoard & 8160 & 1.040 & 2.04 & 0.25 \\ \hline
AES CTR & Kintex Ultrascale & 8160 & 3.876 & 2.58 & 0.31 \\ \hline
\end{tabular}
\end{table*}

\subsection{Experimental Results for Grey Images}

\begin{figure}
\centering
\includegraphics[scale=0.4]{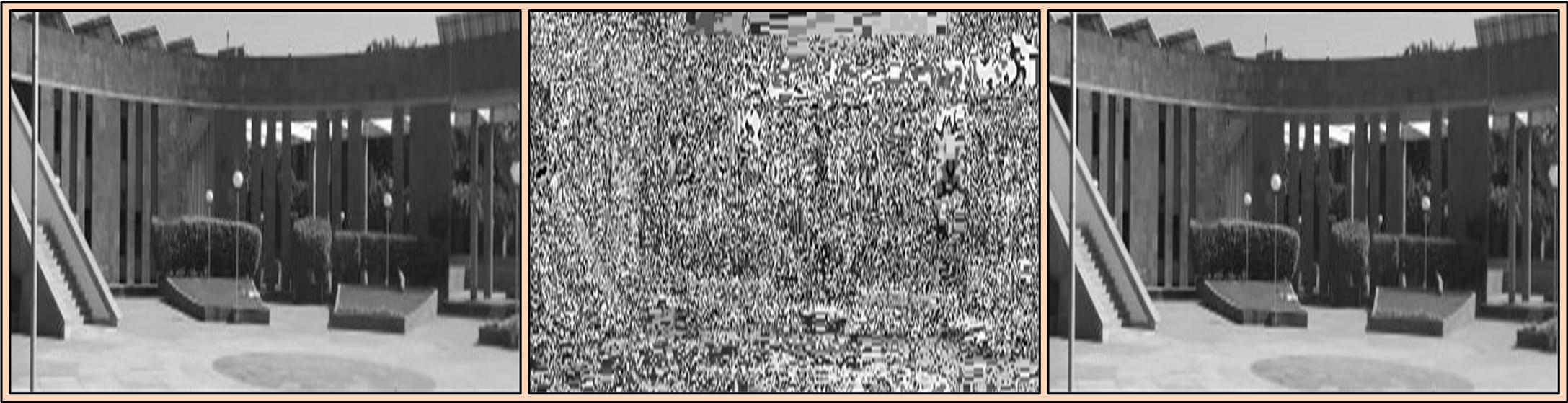}
\caption{Original, encrypted, and decrypted grey image through Hardware Software Co-simulation}
\end{figure}

The mentioned design is an integral part of the XSG based design. Figure 3 shows the encryption and decryption for a grey image by performing the hardware-software co-simulation using XSG. In this image, it can be seen clearly that the encrypted image has patches which means that the encrypted image has the content of the original image. These results are produced with the AES-128 implementation of ECB mode. CTR mode promises several advantages over the ECB mode type of implementation \cite{kane2020security} and it is preferred in high security applications. We also have implemented AES-128 CTR mode and found better results . AES-128 CTR mode is applied to the case study of biomedical images, which is discussed in the next section.

\section{CASE STUDY: HARDWARE SOFTWARE CO-SIMULATION FOR BIOMEDICAL IMAGES (CT Scan Image of Lung Cancer)}
For transferring the secure images through medical edge devices in the IoT environment, the above described Hardware-Software co-design is implemented. For this, the authors have chosen CT scan images of lung cancer \cite{dataset}, to demonstrate the implemented design results. The following subsections present the co-simulation results and security analysis in terms of the histogram, correlation, and Information Entropy analysis of the CT scan image for secure image processing purposes.
\par For performing the secure Hardware Software Co-simulation of Lung Cancer Image, the same model is used as shown in Figure 2. Figure 4 (a) shows the original, encrypted image using ECB \& CTR mode of operation and decrypted images after performing the hardware-software co-simulation using XSG. From the figure, we can conclude that the ability to perform encryption using ECB is very poor, therefore, unpreferable. On the other hand, AES-CTR encryption results are very good for biomedical images.

\begin{figure}
\centering
\includegraphics[scale=0.42]{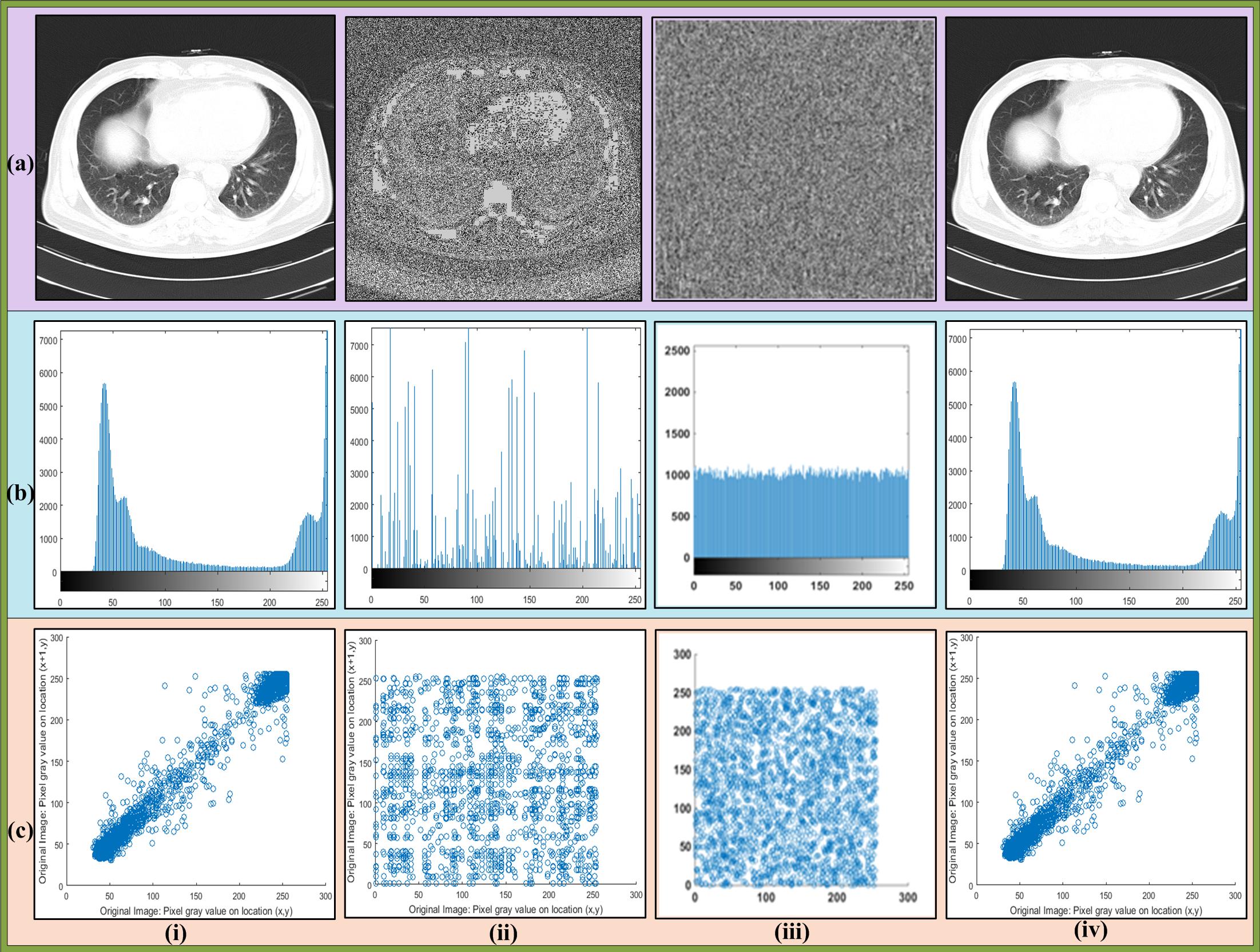}
\caption{CT Scan Input, encrypted using ECB and CTR modes, and decrypted Images, (b) Histogram Analysis, (c) Correlation Analysis of Input, encrypted, and decrypted Images respectively}
\end{figure}
% \caption{CT Scan Input, encrypted using ECB and CTR modes, and decrypted Images, (b) Histogram of Input, encrypted, and decrypted Images, (c) Correlation between pixels of Input, encrypted, and decrypted Images respectively}
\subsection{Security Analysis for the Results of Lung Cancer CT Scan Image: }
This section highlights the security analysis done for the Lung Cancer CT Scan image regarding statistical attacks and key sensitivity analysis.

\subsubsection{Histogram Analysis}
The histogram represents the frequency of occurrence of all gray levels in the image. It describes the distribution of individual pixel values in an image. The histogram analysis of the Lung cancer image, encrypted image using ECB mode, and CTR mode of operation in AES, and the decrypted image is shown in Figure 4 (b). It can be seen that a histogram of the encrypted medical image using AES-CTR is uniform and flat enough to resist statistical attacks (some noise like image data). Therefore, it does not leak any information about the input image.

\subsubsection{Correlation Coefficient Analysis}
In the input image data, there is a strong correlation between adjacent pixels. The correlation distribution of two adjacent pixels in the input image and the encrypted images using both modes of operation is depicted in Figure 4(c). The figure shows that neighboring pixels in the input image are overly correlated, while neighboring pixels in the encrypted image do not correlate. Correlation coefficient results are shown in Table III. It can be seen from the figure and table that the AES-CTR algorithm is not leaking any information about the relationship between the input image and encrypted image, and there is no chance for statistical analysis on medical image processing.

\begin{table}
\centering
\caption{Correlation Coefficient and Information Entropy}
\begin{tabular}{|C {1.8cm}|C {1.2cm}|C {1.7cm}|C {1.5cm}|}
\hline
\textbf{\begin{tabular}[c]{@{}c@{}}Statistical \\ Test\end{tabular}} & \textbf{\begin{tabular}[c]{@{}c@{}}Input \\  Image\end{tabular}} & \textbf{\begin{tabular}[c]{@{}c@{}}Encrypted \\ Image\\ using\\ ECB Mode\end{tabular}} & {\textbf{\begin{tabular}[c]{@{}c@{}}Encrypted \\ Image \\ using \\ CTR Mode\end{tabular}}} \\ \hline
\textbf{\begin{tabular}[c]{@{}c@{}}Horizontal \\ Correlation\end{tabular}} & 0.9929 & -0.0735 & {0.00239} \\ \hline
\textbf{\begin{tabular}[c]{@{}c@{}}Vertical \\ Correlation\end{tabular}} & 0.9825 & -0.0545 & {0.00243} \\ \hline
\textbf{\begin{tabular}[c]{@{}c@{}}Diagonal \\ Correlation\end{tabular}} & 0.9754 & -0.0441 & {0.00210} \\ \hline
\textbf{Entropy} & 6.4406 & 7.6781 & {7.99645} \\ \hline
\end{tabular}
\end{table}

\subsubsection{Information Entropy Analysis}
Information entropy analysis demonstrates the randomness of an information source. It is defined to express the degree of uncertainties in the system. The entropy H(x) of a message source x can be calculated as:
\begin{equation}
\begin{multlined}
H(x)=\sum_{0}^{x-1}p(x_{i })\log_{2}\frac{1}{p(x_{i })}
\end{multlined} 
\end{equation}

The encrypted image’s entropy with 256 (0-255) grey levels should preferably be 8 for a stable cryptosystem. From Table III, we can see that it is 7.6781 for AES-ECB encrypted image and 7.99645 (approximately equal to 8) for AES-CTR encrypted image, which indicates that our AES-CTR system making arduous for attackers to predict pixel values. It is stable enough against entropy attacks. 

\subsubsection{Exhaustive Key Search}
The AES algorithm’s keyspace is $2^{k}$, where k is the number of bits used for the key. We used AES with a 128-bit key size, which means the keyspace is large enough to withstand any brute-force attack. An intruder will have to perform the exhaustive key search $2^{k}$, or $2^{128}$ times, which is virtually impossible to succeed. 

\subsubsection{Key Sensitivity Test}
The key sensitivity analysis guarantees the cryptographic algorithm’s protection. When it comes to key changes, a good crypto algorithm should be extremely sensitive. The CT scan image is encrypted with three different keys to verify this encryption process, i.e., the right key, the same key with a 1-bit difference, and the discrepancy between these two keys. The outcome of using these three different keys, the ciphered images, are presented in Figure 5(b), (c), (d). The decrypted images obtained by applying the right key and 1-bit modified key are shown in Figures 5(e) and 5(f). We can see from these images that the modified key prevents retrieval of the clear image. We have also analyzed the resistance measure of differential attacks by calculating the Number of Pixel Change Rate (NPCR) and Unified Average Changing Intensity (UACI). The NPCR and UACI score obtained to gauge the sensitivity to change in key and their effect in encrypted images are 99.45\% (0.9945) and 33.274\% (0.33274), respectively. Therefore it can be concluded that the implemented design is highly sensitive to key changes.

\begin{figure*}
\centering
\includegraphics[scale=0.7]{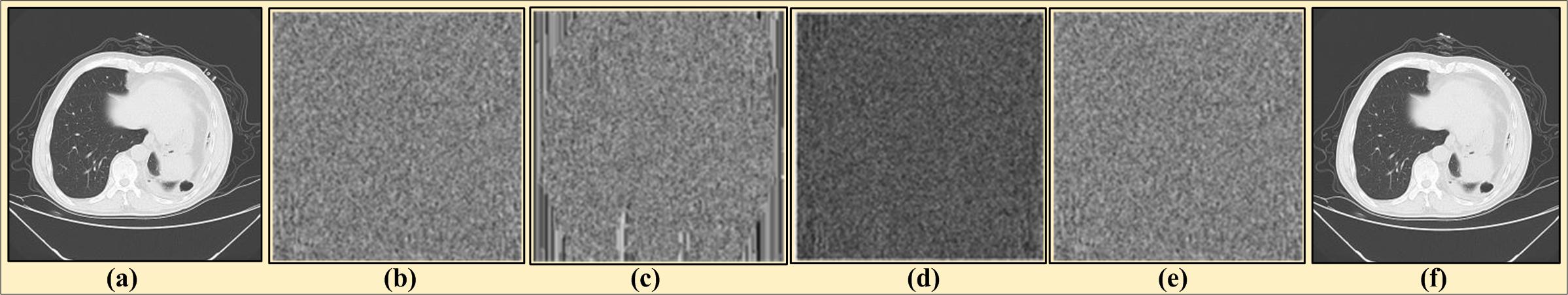}
\caption{Key Sensitivity Test: (a) Original CT scan image of Lung Cancer (b) Cipher image by the right key (c) Cipher image by the 1-bit key change (d) difference between the two ciphered image produced in (b) and (c) (e) Decrypted image by the 1-bit key change (f) Decryption with the right key}
\end{figure*}

\section{Conclusion}

In this article, Hardware-Software Co-Simulation of AES-128 crypto-module is performed using XSG. AES-128 crypto-module is designed using two optimization architectures, i.e., loop unrolled architecture and FSM-based architecture with two modes: ECB and CTR. It is found that FSM-based architecture  for CTR mode gives better performance with high throughput than similar existing implementations. The same approach is also applied to biomedical lung cancer image applications. Security analysis is also performed in terms of the histogram, Correlation coefficient, Information Entropy, and keyspace analysis. Differential attack analysis using key sensitivity test and NPCR \& UACI scores is also performed to prove the robustness of our work.

\bibliographystyle{IEEEtran}
\bibliography{reference}

\section*{ABOUT THE AUTHORS}
\textbf{Kusum Lata} is an Associate Professor with the Department of ECE, The LNMIIT, Jaipur, INDIA. Contact her at kusum@lnmiit.ac.in.
\par \textbf{Surbhi Chhabra} is currently pursuing her Ph.D. in ECE, The LNMIIT, Jaipur, INDIA. Contact her at 16pec004@lnmiit.ac.in.
\par \textbf{Sandeep Saini} is an Assistant Professor with the Department of ECE, The LNMIIT, Jaipur, INDIA. Contact him at sandeep.saini@lnmiit.ac.in.

% In this paper, Hardware-Software Co-Simulation of AES-128 crypto-module is performed using XSG. It is used to encrypt and decrypt the images for IoT Edge devices. \textcolor{blue}{AES-128 crypto-module is designed using two optimization architectures, i.e., loop unrolled architecture and FSM-based architecture with two modes: ECB and CTR.} It is found that FSM-based architecture and AES-CTR give better performance compared to similar existing implementations. A Simulink model with XSG blocks has been developed for simulating the grey and colored images. Then, hardware-software Co-Simulation is performed on Zedboard (XC7Z0201CLG484C) FPGA board, which is one of the latest Xilinx 7 series FPGAs. We have achieved successful encryption and decryption results for both the mode schemes of the AES-128 bit algorithm with high throughput. The same approach is also applied to biomedical image applications where a CT scan image of size 512*512 is taken as a case study. The successful hardware-software co-simulation of encrypted and decrypted CT scan image images can be seen on both the FPGAs. Security analysis is also performed in terms of the histogram, Correlation coefficient analysis, Information Entropy analysis, and keyspace analysis using exhaustive search, key sensitivity test, and NPCR \& UACI scores. 

\end{document}